 \RequirePackage{fixltx2e}
\documentclass[10pt,aps,notitlepage,onecolumn,superscriptaddress,showpacs,nofootinbib,longbibliography,amssymb,showkeys,byrevtex]{revtex4-1}

 \usepackage{amsmath}
\usepackage{amsfonts}
\usepackage{amssymb}
\usepackage{amscd}
\usepackage{accents}
\usepackage{graphicx}
\allowdisplaybreaks[4] 
 
\normalfont
\usepackage{booktabs}
\usepackage{bm}

\renewcommand{\thesection}{\arabic{section}}
\renewcommand{\thesubsection}{\thesection.\arabic{subsection}}
\usepackage{titlesec}
\titleformat{\section}{\centering\large\sffamily\bfseries}{\thesection.}{0.3em}{}
\titleformat{\subsection}{\centering\large\sffamily}{\thesubsection.}{0.3em}{}
\titleformat{\subsubsection}{\centering\sffamily}{\thesubsection.\thesubsubsection.}{0.3em}{}

 \usepackage[colorlinks]{hyperref}
 \usepackage[capitalise]{cleveref}
\usepackage{xcolor}
\definecolor{dark-red}{rgb}{0.6,0.15,0.15}
\definecolor{dark-blue}{rgb}{0.15,0.15,0.8}
\definecolor{medium-blue}{rgb}{0,0,0.6}
\hypersetup{
    colorlinks, linkcolor={dark-red},
    citecolor={dark-blue}, urlcolor={medium-blue}
}

\begin{document}
\title{Wahlquist's metric versus an approximate solution with the same equation of state}

\author{J. E. Cuch\'i}
  \affiliation{Dpto. F\'isica Fundamental, Universidad de Salamanca}
  \thanks{J. E. Cuch\'i: \texttt{jecuchi@usal.es}
E. Ruiz: \texttt{eruiz@usal.es} \\
J. Mart\'in: \texttt{chmm@usal.es}}
 \author{J. Mart\'in}
 \affiliation{Dpto. F\'isica Fundamental, Universidad de Salamanca}
 \author{A.\ Molina}
 \affiliation{Dpt. F\'isica Fonamental, Institut de Ci\`encies del Cosmos, Universitat de Barcelona}
 \thanks{A. Molina: \texttt{alfred.molina@ub.edu}}
 \author{E.\ Ruiz}
 \affiliation{Dpto. F\'isica Fundamental, Universidad de Salamanca}

\begin{abstract}

We compare an approximation of the singularity-free Wahlquist exact solution with a stationary and axisymmetric metric for a rigidly rotating perfect fluid with the equation of state $\mu + 3p= \mu_0$, a sub-case of a global approximate metric obtained recently by some of us. We see that to have a fluid with vanishing twist vector everywhere in Wahlquist's metric the only option is to let its parameter $r_0\rightarrow0$ and using this in the comparison allows us in particular to determine the approximate relation between the angular velocity of the fluid in a set of harmonic coordinates and $r_0$. Through some coordinate changes we manage to make every component of both approximate metrics equal. In this situation, the free constants of our metric take values that happen to be those needed for it to be of Petrov type D, the last condition that this fluid must verify to give rise to the Wahlquist solution.
\end{abstract}
\maketitle

\section{Introduction} 
There are a few exact solutions of the Einstein equations describing the gravitational field inside a
stationary and axisymmetric rotating perfect fluid, the basic candidates to form a  stellar model in General Relativity \cite{senovilla1993contribucion,stephani2003exact}. 
Among them, only  one is known to admit a spheroidal closed surface of zero pressure, the key component to build a stellar model matching the interior (source) spacetime with a suitable asymptotically flat exterior. It is the Wahlquist metric, that describes a rigidly rotating perfect fluid, possesses the energy density-pressure equation of state (EOS) $\mu + 3p =\mu_0\ $ and has Petrov type D \cite{wahlquist1968isf}. Nevertheless, it has been shown in several different ways that it can not correspond to an isolated object nor be matched with an asymptotically flat exterior \cite{wahlquist1968isf,bradley2000wmc,sarnobat2006wes}. Accordingly, General Relativity still lacks  any exact solution that can describe the interior of such stellar model. 


 To find these global models, numerical methods and analytic approximations are therefore the pragmatical way to go. A very influential work for both paths is due to Hartle and Thorne \cite{hartle1967slowly,hartle1968slowly}. They show how to build and match an asymptotically flat vacuum exterior to an interior corresponding to a barotropic and uniformly rotating perfect fluid in slow rotation. The scheme perturbs analytically the non-rotating initial configuration obtaining results up to second order in the slow-rotation parameter. Nevertheless, it usually relies in numerical integration to get them and the matching is not as general as it could. Numerical approximations have been very successful in this field, although some of its most modern and precise codes ---{\ttfamily RNS} \citep{stergioulas1995comparing}, {\ttfamily rotstar} \citep{bonazzola1998numerical,gourgoulhon1999fast}, AKM \citep{ansorg2002highly,ansorg2003highly}, {\ttfamily rotstar-dirac} \citep{lin2006rotating}--- are inspired by the work 
of \citet{ostriker1968rapidly}.  Fully analytic stellar models on the contrary are quite hard to find; this led some of us to introduce a new approximation scheme in \cite{cabezas2006ags,cabezas2007ags} focused in this kind of problem. It is a double approximation. The first one is post-Minkowskian with associated parameter $\lambda$, which is related with the strength of the gravitational field and the second one is a slow rotation 
approximation with parameter $\Omega$, measuring the deformation of the matching surface due to the rotation of the fluid. We have applied this scheme to find an approximate global solution for a fluid with simple barotropic EOS up to order $\lambda^{5/2}$ and $\Omega^3$ in the following cases. We have found solutions for
constant density \cite{cabezas2007ags} and for a polytropic fluid \cite{martin2008crr} with Lichnerowicz matching conditions \citep{lichnerowicz1955theories} and more recently, also for the linear equation of state
 \begin{equation}
\mu + (1-n)p =\mu_0                                                                                                                                                                                          \end{equation}
 with both Darmois-Israel \citep{darmois1966memorial,israel1966singular} and Lichnerowicz matching conditions \cite{Cuchi:2012nmREVTEX} (hereinafter CGMR)\footnote{We had already published some partial previous results in \citep{cuchi2008ags,toni2008ags,cuchi2009matching}}.  In particular ---and contrarily to what happens for anisotropic fluids \cite{Sharma11032007}--- the exact perfect fluid solution in closed form for this EOS is unknown even in the non-rotating case.
 When $n = -2$, this EOS becomes the one of the Wahlquist solution.

It is worth noting that despite the inherent interest of Wahlquist's metric as an exact solution, our metric is in some sense more general even after fixing $n=-2$. This comes from the fact that although Wahlquist's is the most general Petrov type D solution for this EOS, symmetries and motion of the fluid \cite{senovilla1987sap}, our metric can also be of type I. More interestingly, for special values of its constants, our approximate solution, unlike Wahlquist's, can be matched to an asymptotically flat exterior and thus can describe a compact isolated object.

Another problem of Wahlquist's solution is its rotation para\-meter. 
It is written in co-rotating coordinates and lacks of any parameter that can be directly related with the angular velocity of the fluid.
 All that is known is that letting the Wahlquist's $r_0$ parameter go to zero in a limiting procedure (that involves a coordinate change that is singular when $r_0\rightarrow 0$) we get the static spherically symmetric Whittaker solution \cite{wahlquist1968isf,stephani2003exact}. Despite the singular character of this limit, the relation it shows between the Whittaker and Wahlquist metrics seems to be a sound one  since in the slow rotation formalism of \cite{cohen1968further}, which is first order in the rotational parameter, the Whittaker-like and Wahlquist-like slowly rotating metrics coincide \cite{stewart1983rotating,whitman1984comment}.

The question we try to answer in this paper is whether or not we can include an appropriate approximation of the Wahlquist solution in our family of approximate solutions. There are two ways to answer. One of them is asking our $n=-2$ solution to be of  Petrov type
D, since a metric with its characteristics and this Petrov type must belong to the Wahlquist family \cite{senovilla1987sap}. The other way is finding a coordinate change to make them to coincide.

Regarding the first one, we have already verified that our solution can take Petrov type D in \cite{Cuchi:2012nmREVTEX}. Some of its free constants 
 are then fixed and we have found their values do not coincide with the ones they are forced to take when we matched our interior solution with an asymptotically flat exterior solution, as one expects.

The coordinate change way involves writing our solution in
a co-rotating frame, and then making a series expansion of Wahlquist's solution  with $\mu_0$ as post-Minkowskian parameter. When the parameter $\mu_0$ tends to zero the Wahlquist solution becomes
Minkowski's metric, what shows that $\mu_0$ plays an equivalent role to the one of the parameter $\lambda$ in our scheme. This way is more meaningful since, in spite of any result we can get from our approximate metric alone, there is always the question of whether our solution really corresponds to a parametric expansion of an exact metric. Working this way we verify explicitly this correspondence.

In Section 2 we give some notation and definitions used along the paper and 
we write the CGMR interior metric for $n =-2$ and perform the rotation to write our metric
in a co-rotating coordinate system. In Section 3 we give the Wahlquist metric, write it in  spheroidal-like coordinates and then write the approximate
post-Minkowskian Wahlquist metric. Finally, in Section 4 we compare both solutions and
determine the value of our constants and the relation between $r_0$ and the rotation parameter.

\section{The approximate interior metric}
We work within the analytical approximation scheme developed in
 \cite{cabezas2006ags,cabezas2007ags} and \cite{Cuchi:2012nmREVTEX}. It allows to build an approximate stationary and axisymmetric solution of the Einstein's equations for a source spacetime and an asymptotically flat vacuum region around it, although in this paper we put the focus on the interior, source spacetime. This section is devoted to a brief review of its main points in the general formulation. 

 With $\bm{\xi}$ the time-like Killing vector and $\bm{\zeta}$ the space-like closed-orbits Killing vector associated to the stationarity and axisymmetry, let us choose $t$ and $\varphi$ to be coordinates adapted to $\bm{\xi}$ and $\bm{\zeta}$, respectively. Our interior is filled with a perfect fluid without convective motion so we can write its 4-velocity as 
\begin{equation}
\bm{u}=\psi(\bm{\xi}+\omega\bm{\zeta}).                                                                                                                                                                                                                                                                                                                                                                                                                                            \end{equation}
 Here, $\psi$ is a normalization factor and $\omega$ the angular velocity of the fluid in this coordinates. The EOS of the fluid is the $n=-2$ sub-case of the linear one $\mu+(1-n)p=\mu_0$ already studied in \cite{Cuchi:2012nmREVTEX}, i.\,e.
\begin{equation}
 \mu +3p=\mu_0.
\label{wahl_eoswahl}
\end{equation} 
Integrating the Euler equations $\nabla_\alpha T^\alpha_\beta=0$ with this EOS we get the explicit expressions for the mass density $\mu$ and the pressure $p$ in terms of $\psi$ and its value on the $p=0$ surface, $\psi_\Sigma$
\begin{equation}
\begin{aligned}
 p &= \frac{\mu_0}{2}\left(1-\frac{\psi^2_\Sigma}{\psi^2}\right),\\
\mu &= \frac{\mu_0}{2}\left(3\frac{\psi^2_\Sigma}{\psi^2}-1\right).
\end{aligned}
\label{eqpressio1}
\end{equation} 

With this kind of interior, we can choose coordinates $\{r,\,\theta\}$ spanning the 2-surfaces orthogonal to the ones containing $\bm{\xi}$ and $\bm{\zeta}$ \cite{papapetrou1966cgs,kundt1966orthogonal}, then
 we can write the interior --and exterior-- metric with the structure
\begin{align}
\bm{g} &= \gamma_{tt}\,\bm{\omega}^t{\otimes\,}\bm{\omega}^t
+\gamma_{t\varphi}(\bm{\omega}^t{\otimes\,}\bm{\omega}^\varphi+\bm{\omega}^\varphi{\otimes\,}\bm{\omega}^t)+
\gamma_{\varphi\varphi}\,\bm{\omega}^\varphi{\otimes\,}\bm{\omega}^\varphi
\nonumber\\
&\quad +\,\,\gamma_{rr}\,\bm{\omega}^r{\otimes\,}\bm{\omega}^r+
\gamma_{r\theta}(\bm{\omega}^r{\otimes\,}\bm{\omega}^\theta+\bm{\omega}^\theta{\otimes\,}\bm{\omega}^r) 
+\gamma_{\theta\theta}\,\bm{\omega}^\theta{\otimes\,}\bm{\omega}^\theta\,
\label{eqmetrica}
\end{align}
in the associated cobasis $\bm{\omega}^t=dt$, $\bm{\omega}^r=dr$, $\bm{\omega}^\theta=r\,d\theta$, $\bm{\omega}^\varphi=r\sin\theta\,d\varphi$. We will require these coordinates to be spherical-like in the sense that they are associated through the usual relations
\begin{equation}
x=r\sin\theta \cos\varphi,\,\quad y=x=r\sin\theta \sin\varphi,\,\quad z=\cos\theta                                                                       \end{equation} 
to a set $x^\alpha=\{t,\,x,\,y,\,z\}$ of harmonic coordinates ($\square x^\alpha=0$), a particularly relevant gauge choice \cite{choquet1962gravitation,fock1963theory}.

In CGMR we solved the Einstein equations with this coordinate condition using a post-Minkowskian expansion for the metric so that $g_{\alpha\beta}=\eta_{\alpha\beta}+\lambda h^{(1)}_{\alpha\beta}+\lambda^2 h^{(2)}_{\alpha\beta}+\cdots$, with $\eta_{\alpha\beta}$ the flat metric and the approximation parameter 
\begin{equation}
 \lambda=\frac16 \mu_0 r_s^2
\label{wahl_lambdadef}
\end{equation} 
(note that here we work in units where $8\pi G=1$ so this definition is different from the one in CGMR; $r_s$ is the coordinate radius of the surface in the static limit),
using a tensor spherical harmonic expansion truncated using a secondary slow rotation approximation with parameter
\begin{equation}
 \Omega=\omega r_s \lambda^{-1/2}
\end{equation} 
that gives a measure of the deformation of the source.
For $n=-2$, the CGMR interior metric is, up to $\mathcal{O}(\lambda^2,\,\Omega^3)$ \footnote{We have made the notation change $r_0 \rightarrow r_s$ to avoid confusion with the $r_0$ parameter that appears in the usual expressions of the Wahlquist metric. The original CGMR includes $\mathcal{O}(\lambda^{5/2},\,\Omega^3)$ terms as well, but we will not work with them here.}
\begin{align}
   \gamma_{rr}^\text{CGMR}&=1+\lambda\left[m_0-\frac{r^2}{r_s^2}\left(1- m_2\Omega^2P_2\right)\right]\nonumber 
\\
&\quad+\frac{2\lambda^2}{5}\frac{r^2}{r_s^2}\left\{m_0-12S-4 m_0m_2 \Omega^2 P_2- \frac{r^2}{r_s^2}\left[\Omega^2\left(\frac{5}{3}P_2-\frac{8}{7}\right)+\frac{1}{7}\right]\right\}\nonumber
\\
&\quad{}+\mathcal{O}(\lambda^3,\,\Omega^4),
\label{cgmr1}
\\[1ex]
\gamma_{r\theta}^\text{CGMR}&=-\lambda^2\Omega^2 \frac{r^2}{r_s^2}P_2^1 \left[\frac15 m_0m_2+\frac{1}{63}\frac{r^2}{r_s^2}\left(1-6m_2\right)\right]+\mathcal{O}(\lambda^3,\,\Omega^4),
\label{cgmr2}
\\[1ex]
\gamma_{\theta\theta}^\text{CGMR}&=1+ \lambda\left[ m_0 -\frac{r^2}{r_s^2}\left(1- m_2\Omega^2P_2\right)\right]\nonumber
 \\
  &\quad+ \lambda^2\frac{r^2}{r_s^2}\left( -\frac15\left[18S+m_0+2m_0m_2\Omega^2(2P_2-1)\right]\right.\nonumber
 \\
  &\left.\hspace{4em}\quad{}+ \frac17\frac{r^2}{r_s^2}\left\{\frac{8}{5}-\frac{\Omega^2}{3}\left[\frac{m_2}{2}-\frac{134}{15}+\left(\frac{31}{3}+\frac{23m_2}{2}\right) P_2 \right]\right\}\right)\nonumber
\\
&\quad{}+\mathcal{O}(\lambda^3,\,\Omega^4),
\label{cgmr3}
\\[1ex]
\gamma_{\varphi\varphi}^\text{CGMR}&=1+ \lambda\left[ m_0 -\frac{r^2}{r_s^2}\left(1-m_2\Omega^2 P_2\right)\right]+ \lambda^2\frac{r^2}{r_s^2}\left(-\frac15\left(18S+m_0+2m_0m_2\Omega^2\right)\nonumber\right. 
\\
 & \left.\quad{}+ \frac17\frac{r^2}{r_s^2}\left\{\frac{8}{5}+\frac{\Omega^2}{3}\left[\frac{m_2}{2}-\frac{26}{15}+\left(\frac13-\frac{25m_2}{2}\right)P_2 \right] \right\} \right)+\mathcal{O}(\lambda^3,\,\Omega^4),
\label{cgmr4}
\\[1ex]
 \gamma_{t\varphi}^\text{CGMR}&= \lambda^{3/2}\Omega\frac{r}{r_s}\left[\left(j_1-\frac65\frac{r^2}{r_s^2}\right)P_1^1 +j_3\Omega^2\frac{r^2}{r_s^2}P_3^1\right]
+\mathcal{O}(\lambda^{5/2},\,\Omega^5),
\label{cgmr5}
\\[1ex]
\gamma_{tt}^\text{CGMR}&=-1+\lambda\left[m_0-\frac{r^2}{r_s^2}\left(1- m_2\Omega^2 P_2\right)\right]-\lambda^2\frac{r^4}{r_s^4}\left[\frac15\left(1+2\Omega^2\right)-\frac47\Omega^2(1+m_2)P_2\right]\nonumber
\\
&\quad{}+\mathcal{O}(\lambda^3,\,\Omega^4).
\label{cgmr6}
\end{align}
where $P_n^{\,l}$ stands for the associated Legendre polynomials $P_n^{\,l}(\cos\theta)$. With the EOS fixed, the interior in CGMR depends on nine free constants. Two of them, $r_s$ and $\omega$, are part of the approximation parameters $\lambda,\,\Omega$. The other seven are $(m_0,\,m_2,\,j_1,\,j_3,\,a_0,\,a_2,\,b_2)$.  These arise from the harmonic expansion  we use to solve the homogeneous part of the Einstein equations at each order. Accordingly, they are also series expansions in positive powers of $(\lambda,\, \Omega)$. The first four of them are the ones that a Darmois  matching fixes and choosing values for them amounts to choosing a ``particular metric'' from the CGMR family ---although in a strict sense, such particular metric would 
still be a family of metrics because of the free values of $(\lambda,\,\Omega)$---. The last three parametrize changes between the harmonic coordinates used. Here, to simplify  we have taken  these purely gauge constants $a_0=0$, $a_2=0$, $b_2=0$ without losing generality because they are not needed hereafter. The static limit ($\Omega=0$) of CGMR for a certain EOS is characterised with only $(r_s,\,m_0
)$ (see \cref{wahlquist_t:1}).

The constant $S$ is defined as
\begin{align}
\psi_\Sigma=& 1+\lambda  \left(-\frac{1}{2}+\frac{\Omega ^2}{3}+\frac{m_0}{2}\right)
+\mathcal{O}(\lambda^2,\,\Omega^4)\equiv 1+\lambda S +\mathcal{O}(\lambda^2,\,\Omega^4).
\label{wahl_Sdef}
\end{align}
This value $\psi_\Sigma$ comes from the value of $\psi$
\begin{align}
 \psi&=1+\lambda  \left\{-\frac{r^2}{2r_s^2}+\frac{m_0}{2}+\Omega ^2 \left[\frac{r^2}{3r_s^2}+\frac{r^2}{r_s^2} \left(-\frac{1}{3}+\frac{m_2}{2}\right) P_2\right]\right\}\nonumber
\\
&+\lambda ^2 \left(\frac{11 r^4}{40r^4_s}-\frac{3  m_0r^2}{4r^2_s}+\frac{3 m_0^2}{8}+\Omega ^2 \left\{-\frac{7 r^4}{30r_s^4}+\frac{r^2}{r_s^2} \left(-\frac{2 j_1}{3}+\frac{5 m_0}{6}\right)
\right.\right.\nonumber
\\
&+\left.\left.\left[\frac{r^4}{r_s^4} \left(\frac{67}{210}-\frac{13 m_2}{28}\right)+\frac{r^2}{r_s^2} \left(\frac{2 j_1}{3}-\frac{5 m_0}{6}+\frac{3 m_0 m_2}{4}\right)\right] P_2\right\}\vphantom{\frac{11 r_0^4}{40r^4_s}}\right)+\mathcal{O}(\lambda^2,\Omega^2)
\end{align}
on the zero pressure surface 
\begin{equation}
 r(p=0)=r_s\left(1+q \Omega^2P_2\right)
\end{equation} 
where
\begin{equation}
 q=\left(-\frac{1}{3}+\frac{m_2}{2}\right) +\lambda   \left[\frac{1}{21} \left(-1+14 j_1-7 m_0\right)+\frac{3 m_2}{35}\right]+\mathcal{O}(\lambda^2,\Omega^2).
\end{equation} 
%
%
We have not replaced $S$ in the expressions for both brevity and to check the behaviour of $\psi_\Sigma$ when we compare with the parameters in the Wahlquist solution. 
%
\begin{table}[t]
\caption{\label{tab:example}Free constants in the CGMR interior after fixing to zero the pure gauge constants $a_0,\,a_2,$ and $b_2$.}
 \begin{tabular}{ccc}
\toprule
\text{Parameters\ \ \ \ \ \ }&\text{Harmonic expansion constants}&	
{\ \ \ \ \ \ Static limit}\\
\midrule
$\ \mu_0,\,n,\, r_s,\,\omega\qquad$	&$\qquad m_0,\,m_2,\,j_1,\,j_3\qquad$	&
$\qquad\mu_0,\,n,\,r_s,\,m_0\ $\\
\bottomrule
\end{tabular}
\label{wahlquist_t:1}
\end{table}
These expressions for $\psi$ and $\psi_\Sigma$ lead to the following one for the pressure
\begin{align}
 \frac{p}{\mu_0}&=\lambda  \left\{\frac{1}{2}-\frac{r^2}{2r_s^2}+\Omega ^2 \left[-\frac{1}{3}+\frac{r^2}{3r_s^2}+\frac{r^2}{r_s^2} \left(-\frac{1}{3}+\frac{m_2}{2}\right) P_2\right]\right\}+\mathcal{O}(\lambda^2,\,\Omega^4)
\end{align}
from where $\mu(r,\,\theta)$ can be directly obtained using the EOS \labelcref{wahl_eoswahl}. Here we see that, as already happens with Newtonian results for spherical sources \citep{bradley2009quad}, writing $\smash{\mu_0}$ in terms of $\lambda$ with \labelcref{wahl_lambdadef} their lowest orders go as $\mu\sim\lambda$, $p\sim\lambda^2$. 

The range of applicability of CGMR is given by the set of values $(r_s,\mu_0,\,\omega)$. The parameter $\lambda$ will be small whenever $r_s$ or $\smash{\mu_0}$ are small enough. For $\Omega$, small values $\omega$ are in principle required, but the greater $\lambda$ is, the higher $\omega$ can be. This comes from the fact that a strongly gravitationally bounded source deforms much less with rotation than a lightly bounded one.

This solution is apparently less interesting than the Wahlquist exact solution for the same kind of source because it is an approximation. Nevertheless, it is more general in a sense because it is a Petrov type I solution unless  
\begin{equation}
m_2=\frac65+\mathcal{O}(\lambda,\,\Omega^2), \quad j_3=\frac{36}{175}+\mathcal{O}(\lambda,\,\Omega^2),
\label{petrovconD}
\end{equation}
in which case it becomes a Petrov type D solution. It is worth noticing though that when finding the Petrov type of a metric, the more special the algebraic type is, the bigger is the number of conditions to verify. Then, while an approximate metric can satisfy these constraints up to a certain order, it is possible that its higher orders do not. Accordingly, the Petrov type of an approximate metric must be regarded generally as an upper bound to the algebraic speciality of its Weyl tensor (see \citep{Cuchi:2012nmREVTEX}).

Another feature of the CGMR interior is that imposing Darmois-Israel matching conditions \citep{darmois1966memorial,israel1966singular} shows that when
 \begin{align}
m_0&=3+ \lambda  \left(3+ 2\Omega ^2 \right)+\mathcal{O}(\lambda^2,\Omega^4),
\label{int1}\\
m_2&= -1-\frac25\lambda  +\mathcal{O}(\lambda^2,\Omega^2),
\label{int2}
\\
j_1&= 2+\frac{2 \Omega ^2}{3}+\lambda  \left(\frac{44}{5}+\frac{52}{15}\Omega ^2 \right)+\mathcal{O}(\lambda^2,\Omega^4),
\label{int3}\\
j_3&= -\frac{2}{7}-\frac{296}{245}\lambda  +\mathcal{O}(\lambda^2,\Omega^2)
\label{int4}
 \end{align}
the interior can be matched with an asymptotically flat vacuum exterior \cite{Cuchi:2012nmREVTEX}.
Additionally, in our solution the parameter $\omega={u^\varphi}/{u^t}$ is the angular velocity of the fluid with respect to our harmonic coordinate frame and its vanishing leads to a static solution (i.\,e. $\smash{\gamma_{t\varphi}}=0$). There is no parameter in Wahlquist's metric with these two features.

If we want to compare this approximate solution with Wahlquist's metric we have to start finding their expressions in the same coordinates.  The first problem is that the Wahlquist metric is written in a co-rotating coordinate system and CGMR is not, so first we must choose between the two kinds of coordinates. Changing the CGMR interior to a co-rotating system is straightforward doing 
\begin{equation}
\varphi\rightarrow \varphi+\frac{\lambda^{1/2}\Omega}{r_s}t ,\quad t\rightarrow t
\label{stopchange}
\end{equation}
and then in the co-rotating system the metric components are:
\begin{align}
\gamma_{tt}^\text{CGMR}&=-1+\lambda\left\{m_0+\frac{r^2}{r_s^2}\left[-1+\frac13\Omega^2\left(2+(3m_2-2)P_2\vphantom{A^A}\right)\right]\right\}\nonumber \\
&\quad{}+\lambda^2\frac{r^2}{r_s^2}\left\{\frac23\Omega^2(2j_1-m_0)(P_2-1)-
\frac15\frac{r^2}{r_s^2}\left[1-\frac23\Omega^2\left(4+\frac{1}{7}(30m_2-19)P_2\right)\right]\right\}\,\nonumber\\
&\quad{}+\mathcal{O}(\lambda^3,\,\Omega^4),
\label{cgmr7}
\\
\gamma_{t\varphi}^\text{CGMR}&= -\Omega \lambda^{1/2}\frac{r}{r_s}\left(
P_1^1+\lambda\left\{\left[m_0-j_1+\frac15\frac{r^2}{r_s^2}\left(1-\Omega^2 m_2\right)\right]P_1^1\right.\right. \nonumber
\\
 &\left.\left.\!\!\quad{}- \Omega^2\frac{r^2}{r_s^2}\left(j_3-\frac{m_2}{5}\right)P_3^1\right\}
\right)
+\mathcal{O}(\lambda^{5/2},\,\Omega^5)
\label{cgmr8}
\end{align}
and the other components remain unchanged. Let us remark that the $\smash{\gamma_{t\varphi}}$ component is now of order $\smash{\lambda^{1/2}}$ instead of the order $\lambda^{3/2}$ it was in the original coordinates (see \cite{Cuchi:2012nmREVTEX} for some comments).

\section{ The  Wahlquist metric}

The next steps in the comparison are, using the singularity free Wahlquist metric,  first expand it in the appropriate approximation parameters and then make coordinate changes to reduce it to a particular case of the CGMR interior.

The singularity free Wahlquist metric reads  \cite{wahlquist1968isf,stephani2003exact}
\footnote{Here $\xi$ is a coordinate not to be mistaken with any quantity related with the stationary Killing vector $\bm{\xi}$ of CGMR}
\begin{align}
&ds^2=-f(dt+A\,d\varphi)^2 +\notag
\\[1ex]
&\hspace{3em} r_0^2(\xi^2+\eta^2)\!\left[\frac{c^2 h_1 h_2}{h_1-h_2}\,d\varphi^2+\frac{d\xi^2}{(1-k^2\xi^2)h_1}+ \frac{d\eta^2}{(1+k^2\eta^2)h_2}\right]
\label{wahlquist}
\end{align}
where
\begin{align}
&f(\xi,\eta)=\frac{h_1-h_2}{\xi^2+\eta^2} ,\quad A= c\,r_0\!\left(\frac{\xi^2 h_2+\eta^2h_1}{h_1 -h_2}-\eta_0^2\right)
\\[1ex]
&h_1(\xi) = 1+\xi^2 +\frac{\xi}{b^2}\left[\xi-\frac1{k}(1-k^2\xi^2)^{1/2}\arcsin(k\,\xi)\right]
\\[1ex]
&h_2(\eta) = 1-\eta^2 -\frac{\eta}{b^2}\left[\eta-\frac1{k}(1+k^2\eta^2)^{1/2}\text{arcsinh}(k\,\eta)\right]
\end{align}
and
\begin{equation}
k^2\equiv \frac12\,\mu_0\,r_0^2 b^2\,.
\label{wahl_eqk}
\end{equation} 
Here 
$\mu_0, b,r_0$ are free constants and $\eta_0$ and $c$ are related with the behaviour of the solution on the axis. The symmetry axis is located at $\eta=\eta_0$ where
\begin{equation}
h_2(\eta_0)=0\,,
\label{wahl_eqeta0}
\end{equation} 
and to satisfy the regularity condition of axisymmetry, $c$ must be
\begin{equation}
\frac1{c}=\frac12 (1+k^2\eta_0^2)^{1/2}\left.\frac{dh_2}{d\eta}\right|_{\eta=\eta_0}\,.
\label{wahl_eqc}
\end{equation}
Therefore $\eta_0$ and $c$ become functions of the constants $\mu_0,\, r_0$ and $b$, which thus characterise completely the singularity free Wahlquist's solution.
It is generated by a perfect fluid with 4-velocity
\begin{equation}
\bm{u} = f^{-1/2}\,\partial_t \qquad (g_{\alpha\beta}u^\alpha u^\beta=-1)\,,
\end{equation} 
and with energy density and pressure are given by
\begin{align}
&\mu =\frac12\mu_0(3b^2 f-1)
\\[1ex]
&p =\frac12\mu_0(1-b^2 f)
\label{pressure_density}
\end{align} 
where we can see now more clearly that the constants $b$ and $\smash{\mu_0}$ are the values of  the normalization factor $f^{-1/2}$  and the energy density on the matching surface of zero pressure (see also \eqref{eqpressio1}). 

Regarding rotation in Wahlquist's solution, the full expression of  the module of its twist vector $\bm{\varpi}^\text{W}(\eta,\,\xi)$ can be found in \cite{wahlquist1968isf} and its 
value  at  $(\eta=0,\,\xi=0)$ is
\begin{align}
\varpi^\text{W}(0,\,0)&=\frac13 \mu_0 r_0.
\label{twist2}
\end{align} 
 We can also get a static limit for it ---Whittaker's metric \cite{whittaker1968interior}--- making the change 
\begin{equation}
\{\xi,\,\eta\}\rightarrow\{{R},{\chi}\}:\ \{\xi=\frac{{R}}{r_0},\eta=\cos{\chi}\}
\label{staticlimit} 
\end{equation}
 and letting $r_0$ go to zero \cite{wahlquist1968isf} although it must be noted that this coordinate change is singular when $r_0=0$.

Expression 
\labelcref{twist2}
 and the limiting procedure suggest a relation between $r_0$ and the rotation of the fluid. It is actually the case since  
\begin{equation}
 \lim_{r_0\rightarrow0}\varpi^\text{W}=0
\end{equation} 
 everywhere so $r_0\rightarrow0$ implies vanishing rotation and should lead to a static spacetime. Nevertheless, 
it must be done through the limiting procedure \labelcref{staticlimit}. It is also worth noticing that the only other parameter choice capable of giving $\varpi^\text{W}=0$ everywhere is $\mu_0=0$ but it gives an empty interior. 


 The parameters $\{r_0,\,\mu_0\}$ will be the natural choice for us to make the formal expansions ---they do not need to be small at all--- of the Wahlquist metric  if we want to compare with the post-Minkowskian and slow rotation expansions of CGMR, but first we must find the change to spherical-like coordinates.

\subsection{The Wahlquist metric written in spherical--like coordinates}

Our approximate metric \labelcref{cgmr1,cgmr2,cgmr3,cgmr4,cgmr7,cgmr8} is written in ``standard" 
spherical coordinates ---in the sense that when $\lambda=0$ the metric becomes the Minkowski metric in standard spherical coordinates--- so we need to find a consistent way to write the Wahlquist metric in a set of coordinates as close to ours as possible to begin with.

In this regard we note first that if we put $ \mu_0 = 0 $ in the Wahlquist metric \labelcref{wahlquist} we obtain the Minkowski metric in oblate spheroidal coordinates $\{\xi, \eta \}$, whose coordinate lines are oblate confocal ellipses and confocal  orthogonal  hyperbolas. From these coordinates it is easy to go to Kepler coordinates $\{R, \chi \}$ changing $ \xi = R/r_0, \eta = \cos \chi $, where $R$ represents the semi-minor axis of the ellipses, $\chi$ the Kepler eccentric polar angle and $r_0$ the focal length. Finally we get standard spherical coordinates $\{r, \theta \}$ by changing
\begin{equation}
\sqrt{R ^ 2 + r_0 ^ 2} \sin \chi = r \sin \theta,\quad R \cos \chi = r \cos \theta.                                                                                
\label{KeplerAndPolar}
\end{equation} 

 Moreover, the limiting procedure \labelcref{staticlimit} from Wahlquist's solution to its static limit (the Whittaker metric) has a similar form, in this case leading to Kepler-like coordinates.

\begin{figure}[b]
\begin{tabular}{cc}
 \includegraphics[width=16em]{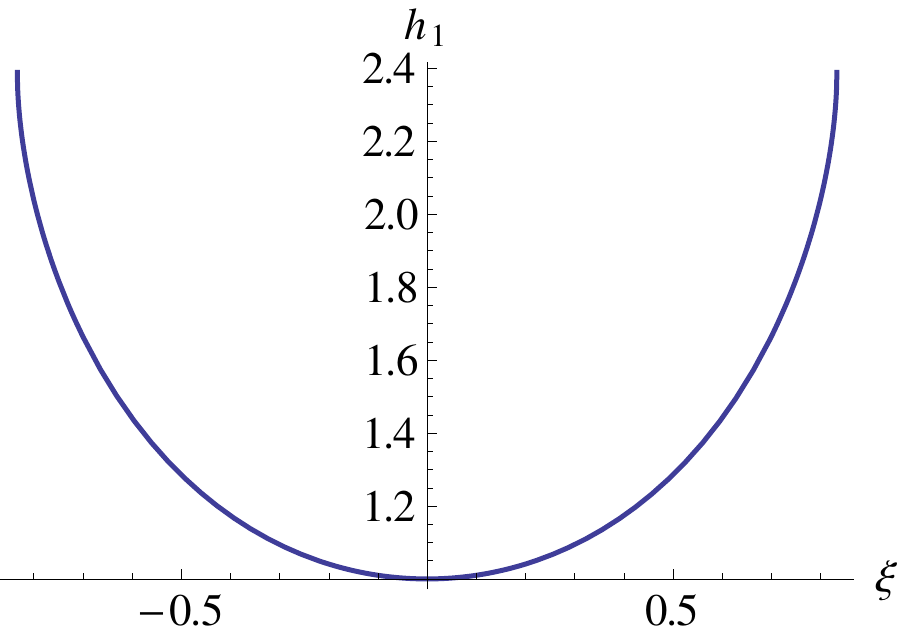}&\includegraphics[width=16em]{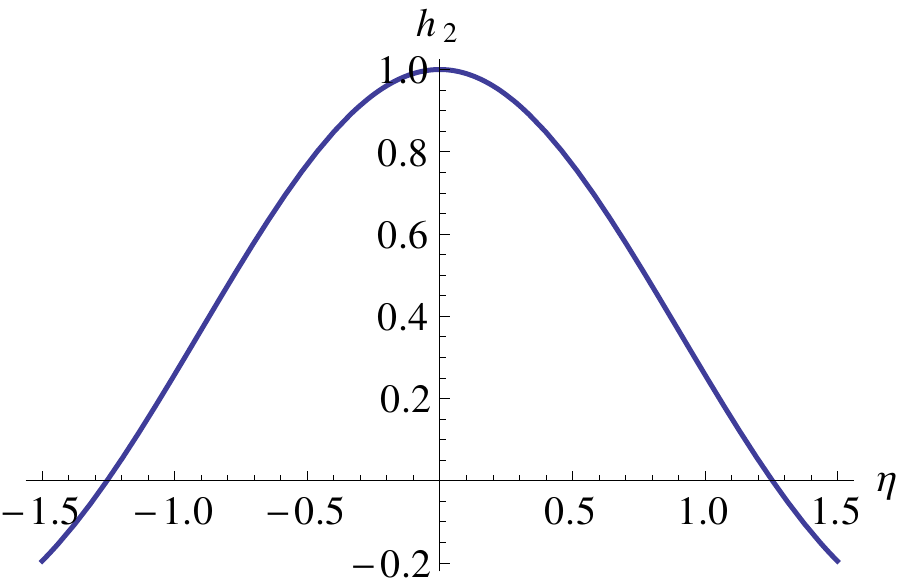}
\end{tabular}
\caption{Behaviour of the $h_1(\xi)$ and $h_2(\eta)$ functions for $k=1.2,\,b=1$ and $k=1.248,\,b=1$, respectively}
\label{figWahl1}
\end{figure}

These considerations suggest to look for a change of coordinates in the Wahlquist metric (prior to any limit) so that the new coordinates ``directly represent''  spheroidal--like  coordinates.  We use a heuristic approach here and start plotting the graphs of  $h_1(\xi)$ and $h_2(\eta)$ (\cref{figWahl1}.)
%
We can see that these curves have the appearance  of a hyperbolic cosine and a squared sine for some values of $b$ and $k$, respectively. Taking this into account we write as an educated guess

\begin{equation}
\{\xi,\eta\}\rightarrow \{{R},\,{\chi}\}:\quad
\left\{\begin{aligned}
&h_1(\xi) =1+ \frac{{R}^2}{r_0^2}= 1+ \frac{R_1^2}{r_0^2}
\\
& 1-h_2(\eta) = \cos^2{\chi}=\frac{R_2^2}{r_0^2}\,.
\label{canvi1}
\end{aligned}\right.
\end{equation}
where we introduce $R_1,\, R_2$ just to simplify calculations later.
Let us write now  the two dimensional metric spanned by $\{\xi,\eta\}$
\begin{equation}
 d\Sigma^2=A d\xi^2+B d\eta^2
\label{sigmametric}
\end{equation}
 in terms of $\{R_1,\,R_2\}$. Since
\begin{equation}
d\xi=\dfrac{dh_1}{dh_1/d\xi}\quad \text{and}\quad d\eta=\dfrac{dh_2}{dh_2/d\eta}\,,
\end{equation} 
taking $h_1,\,h_2$ as functions of $R_1$ and $R_2$
\begin{equation}
dh_1=\frac{2R_1}{r_0^2} dR_1\,,
\quad
-dh_2=\frac{2R_2}{r_0^2}dR_2\,,
\end{equation}
we get 
\begin{equation}
d\Sigma^2=m_{11}\left(\frac{2R_1}{r_0^2} dR_1\right)^2+m_{22}\left(\frac{2R_2}{r_0^2}dR_2\right)^2 ,
\end{equation}
where 
\begin{equation}
 m_{11}=\frac{A}{(dh_1/d\xi)^2},\quad m_{22}=\frac{B}{(dh_2/d\eta)^2}.
\end{equation} 

Now let us do another coordinate change to a kind of spherical coordinates $\{r$, $\theta\}$ using the previous relations \cref{canvi1,KeplerAndPolar}. Then, $R_1$ and $R_2$ are the following functions of $r$ and $\theta$ 
\begin{equation}
\{R_1,R_2\}\rightarrow \{r,\theta\}:\quad
\left\{\begin{aligned}
R_1&=\sqrt{\frac12\left(r^2 - r_0^2 + \sqrt{(r^2- r_0^2)^2 + 4r^2r_0^2\cos^2\theta}\right)}\,, \\[1ex] 
R_2&=\sqrt{\frac12\left(r_0^2 - r^2 + \sqrt{(r^2- r_0^2)^2 + 4r^2r_0^2\cos^2\theta}\right)}\,,
\end{aligned}
\right.
\end{equation} 
and if we define the function
\begin{equation*}
F=(r^2-r_0^2)^2+4r^2r_0^2\cos^2\theta\,,                                                    \end{equation*}
we have that the metric \labelcref{sigmametric} in $\{r,\,\theta\}$ coordinates
$$
d\Sigma^2=g_{rr}dr^2+2g_{r\theta}drd\theta+g_{\theta\theta}d\theta^2
$$
has the coefficients
\begin{align}
g_{rr}&=\frac{2r^2}{r_0^4F}\left[\sqrt{F}(r^2-r_0^2+2r_0^2\cos^2\theta)(m_{11}-m_{22})\right.\nonumber
\\
&\left.\hspace{9em}{}+(F-2r_0^4\sin^2\theta\cos^2\theta)(m_{11}+m_{22})\right],
\\
g_{r\theta}&=-\sin\theta\cos\theta\frac{2r^3}{r_0^2F}\left[\sqrt{F}\left(m_{11}-m_{22})\right.\right.\nonumber
\\
&\left.\left.\hspace{9em}{}+(r^2-r_0^2+2r_0^2\cos^2\theta\right)\left(m_{11}+m_{22}\right)\right],\\
g_{\theta\theta}&=\sin^2\theta\cos^2\theta\frac{4r^4}{F}(m_{11}+m_{22}).
\end{align}
Only the terms $m_{11}+m_{22}$ and $m_{11}-m_{22}$ depend  on both $\{\mu_0,\,r_0\}$; the remaining terms depend on $r_0$ alone.

 Now we are going to write the full metric in terms of $\{r,\theta\}$; notice that the inversion can only be approximately done (in a series of $\mu_0$).
First, we determine $\eta_0$ up to order $\mu_0^2\,$, and then $c$ to the same order. This last series depend on $b^2$, so before that must 
 determine how $b$ depends on $\mu_0$. We recall that in the $\mu_0=0$ limit the Wahlquist metric becomes Minkowski's metric written in oblate spheroidal coordinates so
\begin{equation}
 \lim_{\mu_0\rightarrow0}f=1,
\end{equation} 
 and hence, since $\left.b=f^{-1/2}\right|_{p=0}$, its series expansion must begin as  
$b^2=1+\mathcal{O}(\mu_0)$. Besides, since \cref{pressure_density} takes the form
\begin{equation}
 p=\frac12\mu_0\left\{1-b^2[1+\mathcal{O}(\mu_0)]\right\}\,,
\end{equation}
the expansion of $b$ makes the pressure start with $p\sim\mu_0^2$, behaving like in CGMR. Accordingly, we are going to use
\begin{equation}
b^2 = 1+\frac13\mu_0\sigma_1+\mu_0^2\sigma_2+\mathcal{O}(\mu_0^3)
\label{b-aprox}
\end{equation}
where $\sigma_1$ and $\sigma_2$ are two new constants introduced merely for calculation convenience. Inserting it into \cref{wahl_eqk,wahl_eqeta0,wahl_eqc}, we obtain for the constants $\eta_0$ and $c$ up to $\mathcal{O}(\mu_0^3)$
\begin{align}
\eta_0&= 1+\frac1{12}\mu_0r_0^2\left\{1+\frac{1}{120}\mu_0r_0^2\left[11+\mu_0\left(\frac{73}{28}r_0^2-8\sigma_1\right)\right]\right\}+
\mathcal{O}(\mu_0^4),\\[2ex]
c &= -1+\frac1{12} r_0^2 \mu_0^2\left[\sigma_1-\frac{r_0^2}{3}+\mu_0\left(3\sigma_2-\frac{r_0^4}{30}\right)\right]+\mathcal{O}(\mu_0^4).
\end{align}
Next, we invert the change of coordinates \cref{canvi1}, which gives
\begin{align}
\xi^2 &=  \frac{R_1^2}{r_0^2}\left(1-\frac16 \mu_0 R_1^2\left\{1-\frac{1}{15}\mu_0 R_1^2\left[2-\mu_0\left(\sigma_1+\frac{37}{84}R_1^2\right)\right]\right\}\right)+\mathcal{O}(\mu_0^4)
\\[2ex]
\eta^2 &=  \frac{R_2^2}{r_0^2}\left(1+\frac16 \mu_0 R_2^2\left\{1+\frac{1}{15}\mu_0 R_2^2\left[2-\mu_0\left(\sigma_1-\frac{37}{84}R_2^2\right)\right]\right\}\right)+\mathcal{O}(\mu_0^4)
\end{align}
And finally, by doing the coordinate change $\{R_1,R_2\}\rightarrow \{r,\theta\}$ 
we obtain the metric coefficients up to $\mathcal{O}(\mu_0^3)$ in the spherical--like coordinates desired
\begin{align}
\gamma_{rr}^\text{W}&= 1+\frac{\mu_0}{6}(r_0^2-r^2)+\frac{\mu_0^2}{6}\left[\sigma_1(r^2-r_0^2\sin^2\theta)+r_0^2\left(\frac{4r^2}{5}-\frac{r_0^2}{3}\right)\cos^2\theta\right.\nonumber
\\
&\left.\!\quad{}+\frac{7}{15}(r_0^2-r^2)^2\right]+\frac{\mu_0^3}{90}\left\{\vphantom{\frac{A^A}{A}}\frac{\sigma_1}{2}\left[ r_0^2\left(5r_0^2-7r^2\right)\cos^2\theta-7(r_0^2-r^2)^2\right]
\right.\nonumber
\\
&\quad{}+45\sigma_2(r^2-r_0^2\sin^2\theta)+\frac{r_0^2}{21}(r_0^2-r^2)(85r^2-28r_0^2)\cos^2\theta \nonumber
\\
& \left.\!\quad{}+\frac{149}{84}(r_0^2-r^2)^3-\frac{r_0^4}{2}r^2\cos^4\theta
\right\}
\label{appWahlquist}
\\[1ex]
 \gamma_{\theta\theta}^\text{W}&=1+\frac{\mu_0}{6}(r_0^2-r^2)+\frac{\mu_0^2}{9}\left[r_0^2\left(\frac{r^2}{5} +\frac{r_0^2}{2}-\frac{3}{2}\sigma_1\right)\cos^2\theta+\frac1{5}(r^2-r_0^2)^2 \right] \nonumber
\\
&\quad{}+\frac{\mu_0^3}{180}\left\{\vphantom{\frac{A^A}{A}}
\sigma_1\left[r_0^2(3 r^2-5 r_0^2)\cos^2\theta -2(r_0^2-r^2)^2\right]-90 \sigma_2r_0^2\cos^2\theta \right.\nonumber
\\
&\left.\quad\! {}+\frac{r_0^2}{21} (r_0^2-r^2)(37r^2+56r_0^2)\cos^2\theta+r_0^4 r^2  \cos^4\theta+\frac{37}{42}(r_0^2-r^2)^3
\right\}
\\[1ex]
 \gamma_{r\theta}^\text{W}&=\frac{\mu_0^2r_0^2}{18}\sin\theta\cos\theta\left\{r_0^2-r^2-3 \sigma_1\right.\nonumber
\\
&\left.\quad{}-\frac{\mu_0}{10}\left[\frac{}{}5 \sigma_1 (r_0^2-r^2)+90 \sigma_2-r_0^2r^2\cos\theta^2-\frac{8}{3}(r_0^2-r^2)^2\right]
\right\}
\\[1ex]
\gamma_{ \varphi \varphi}^\text{W}&=1+\frac{\mu_0}{6} ( r_0^2-r^2)  -\frac{\mu_0^2}{6}  \left\{ r_0^2\left[\sigma_1-\frac{1}{15}\left(7  r_0^2-\frac{13}{2}r^2\right)\right]-\frac{3}{10}r_0^2 r^2 \cos^2\theta \right.\nonumber
\\
&\left.\quad{}- \frac{2}{15}  r^4\right \}+\frac{\mu_0^3}{90}  \left\{\sigma_1\left[\frac{r_0^2}{2}(9r^2-7r_0^2)-r_0^2r^2\cos^2\theta-r^4\right]-45r_0^2\sigma_2\right.\nonumber
\\
&\left.\quad{}+\frac{149}{84}r_0^6-\frac{43}{14}r_0^4r^2+\frac{11}{7}r_0^2r^4-\frac{37}{84}r^6-\frac{1}{84}r_0^2r^2(95r^2-151r_0^2)\cos^2\theta
\right\}
\\[1ex]
\gamma_{t\varphi}^\text{W}&= -\frac{\mu_0r_0}{6}r  \sin\theta  \left\{1  +\frac{\mu_0}{30}    (r^2+3  r_0^2)  +\frac{\mu_0^2}{15}\left[  \sigma_1\left(r^2-\frac{13}{4}r_0^2\right)\right.\right.\nonumber
\label{wahl_wahltphi}
\\
&\left.\left.\!\quad{}+\frac{r_0^2}{84}(97r_0^2-41r^2)+\frac{4}{21}r^4+\frac{3}{28}r_0^2r^2\cos^2\theta\right]\right\}
\\[1ex]
\gamma_{tt}^\text{W}&=-1+\frac{\mu_0}{6}  (r_0^2-r^2) -\frac{\mu_0^2}{180}\left\{(r_0^2-r^2)^2-4  r^2  r_0^2  \cos^2\theta\right\}+\frac{\mu_0^3}{90}\left\{(r_0^2-r^2)\frac{}{}\times\right.\nonumber
\\
&\left. \!\quad{}\times \left[\frac4{21}  (r_0^2-r^2)^2+\frac3{14}  r^2  r_0^2  \cos^2\theta\right] -\sigma_1  \left[(r^2-r_0^2)^2+r^2  r_0^2  \cos^2\theta\right]  \right\}
\label{wahl_wahltt}
\end{align}

\section{Comparing  the approximate Wahlquist solution  with the CGMR  solution in co-rotating coordinates }

Now we face the problem of identification of the parameters and to perform the final adjustments of coordinates needed to make every term in Wahlquist's metric and the CGMR interior equal.

 To get an idea of the problems arising, we analyze first the static limit. Using \cref{wahl_lambdadef} and making $r_0=0$ in \cref{appWahlquist} we obtain the expression for  the $\gamma_{rr}$ coefficient of the  static metric
\begin{equation}
\gamma_{rr}^{\text{W}}(r_0=0)= 1-\frac{r^2 \lambda }{{r_s}^2}+\frac{2 r^2 \lambda ^2
 \left(7 r^2+15 \sigma_1 \right)}{5 {r_s}^4}+\mathcal{O}(\lambda^3)
\end{equation}
and upon comparison with the corresponding static limit of CGMR
\begin{equation}
\gamma_{rr}^{\text{CGMR}}(\Omega=0)=1+{m_0}
 \lambda -\frac{r^2 \lambda }{{r_s}^2}-\frac{2 r^4 \lambda ^2}{35 {r_s}^4}+\frac{2 {m_0} r^2 \lambda ^2}{5 {r_s}^2}-\frac{24 r^2 S
 \lambda ^2}{5 {r_s}^2}+\mathcal{O}(\lambda^3)
\end{equation}
we can see that there are discrepancies among  $r^4$ terms in the sense that they can not be made equal adjusting parameters. To some extent this was to be expected since CGMR was written in coordinates associated to harmonic ones and no such a condition has been imposed on the Wahlquist metric. In this particular case, the two metrics can be rendered exactly equal with a change of radial coordinate in $\gamma_{\alpha\beta}^\text{W}(r_0=0)$
\begin{equation}
 r\to r' \left[1+\left(-\frac{2 r'^4}{7 r_s^4}-\frac{9 r'^2 S}{5 r_s^2}\right) \lambda ^2\right]+\mathcal{O}(\lambda^3)
\end{equation} 
and making $m_0=0,\, \sigma_1 = r_s^2 S$.
\subsection{Adjusting parameters}
We go back now to the non--static case.  If we compare the lowest order term in $\smash{g_{t \varphi }}$ and $\smash{g_{tt}}$ of both solutions (\cref{cgmr7,cgmr8,wahl_wahltt,wahl_wahltphi}) we can see that he relation between $\lambda$ and $\mu_0$ is \labelcref{wahl_lambdadef} as expected and the constant $\Omega$ of the CGMR solution must be related with the $r_0$ constant of the Wahlquist metric as follows
\begin{equation}
 r_0=-\frac{\kappa\, r_s\Omega}{\lambda^{1/2}}\label{constsubst}\end{equation} 
with $\kappa$ a factor to be determined later on. If we perform this identification we get to a new difficulty because Wahlquist's solution has  $\lambda$-free terms with  $\Omega$ dependence.  These terms appear associated with powers of $\mu_0r_0^2$ \big[or $\kappa^2\Omega^2$ using \cref{constsubst}\big]. This is not possible in our self--gravitating solution building scheme.
This issue can be solved using the remaining freedom in time scale and $\{r, \theta\}$ coordinates. The changes we can do are\footnote{Here $R$ is a totally new coordinate not to be mistaken with the one used in \cref{KeplerAndPolar}} 
\begin{align}
&t=T\left(1+\mu_0 F_1 +\mu_0^2 F_2+\cdots\right)
\\[1ex]
&\!\!\begin{aligned} 
r&=R\big[1+\mu_0 G_1(R,\Theta)+\mu_0^2 G_2(R,\Theta)+\cdots\big]
\\[1ex]
\theta&=\Theta+\mu_0 \sin\Theta \big[H_1(R,\Theta)+\mu_0 H_2(R,\Theta)+\cdots\big]\,,\label{change}\end{aligned}
\end{align}
with $F_i$ constants depending on the parameters and $G_i,\, H_i$ undetermined functions. Imposing vanishing of these unwanted terms, we get the time scale change
\begin{align}
t& = T\left\{1+\frac{\mu_0r_0^2}{12}\left[1+\frac{11\mu_0r_0^2}{120}\left(1+\frac{73\mu_0r_0^2}{308}\right)\right]\right\}+\mathcal{O}(\mu_0^4)\label{ttrans}
\\
\intertext{and the $\{r,\,\theta\}$ changes}
r&=R\left\{1-\frac{\mu_0r_0^2}{12}\left(1+\frac{\mu_0r_0^2}{3}\left[\frac{41}{40}-\cos^2\Theta+\frac{\mu_0r_0^2}{60}\left(\frac{191}{56}-\cos^2\Theta\right)\right]\right)\right\}+\mathcal{O}(\mu_0^4),
\label{wahl_rtrans}
\\ 
\theta &= \Theta-\frac{\mu_0^2r_0^4}{36}\sin\Theta\cos\Theta\left(1+\frac{\mu_0r_0^2}{10}\right)+\mathcal{O}(\mu_0^4).
\label{wahl_thetatrans}
\end{align}
Note that the symmetry axis for the old coordinates is located at $\theta=0,\,\pi$ and due to the presence of the $\sin\Theta$ it remains at $\Theta=0,\pi$. We will maintain this condition for all the coordinate changes of the $\theta$ coordinate.

Now, we introduce these changes in our last expression of Wahl\-quist metric 
obtaining up to $\mathcal{O}(\mu_0^3)$
\begin{align}
\gamma_{RR}^\text{W}&=1- \frac{\mu_0}{6} R^2+ \mu_0^2 \left\{\frac{7}{90} R^4+\frac{\sigma_1}{6}  R^2+ r_0^2 \left[\frac{R^2}{10}\left(\frac43\cos^2\Theta-1\right)-\frac{\sigma_1}{6}  \sin^2\Theta\right]\right\} \nonumber
\\
&\quad{}+\mu_0^3 r_0^2\left[\frac{17}{42} R^4\left(\frac{1}{20}-   \frac{1}{9}\cos^2\Theta\right)-\frac{\sigma_2}{2}\sin^2\Theta+\frac{\sigma_1}{45} R^2\left(1-\frac{7}{4} \cos^2\Theta\right)  \right],
\\[1ex]
\gamma_{R\Theta}^\text{W}&=- \sin\Theta  \cos\Theta  \frac{\mu_0^2 r_0^2}{6}\left[ \sigma_1+\frac{1}{3} R^2+ \mu_0 \left(3  \sigma_2-\frac{\sigma_1}{6}   R^2-\frac{4}{45} R^4\right)\right],
\\[1ex]
\gamma_{\Theta\Theta}^\text{W}&=1-\frac{\mu_0}{6}R^2+\frac{1}{45}\mu_0^2\left\{ R^4  +r_0^2\left[  R^2 \left( \cos\Theta^2+\frac12\right)-\frac{15}{2}\sigma_1  \cos^2\Theta \right]\right\}  \nonumber
\\
&\quad{}+\frac{\mu_0^3}{2}r_0^2\left[ \frac{\sigma_1}{90} R^2\left(  4+3\cos\Theta^2\right)+\frac{1}{140} R^4\left(1-\frac{74}{27} \cos^2\Theta\right)-\sigma_2  \cos^2\Theta\right],
\\[1ex]
\gamma_{\varphi\varphi}^\text{W}&=1-\frac{\mu_0}{6}R^2+\frac{\mu_0^2}{3}\left\{ \frac{1}{15}R^4 +\frac{1}{2}r_0^2\left[\frac{R^2}{10}(3\cos\Theta^2-1)-  \sigma_1\right]\right\}\nonumber
\\
&\quad{}+ \mu_0^3r_0^2\left[\frac{\sigma_1}{90} R^2\left(\frac{9}{2}-\cos^2\Theta\right)+\frac{R^4}{63}\left(\frac25-\frac{19}{24}\cos^2\Theta\right)-\frac{\sigma_2}{2}\right],
\\[1ex]
\gamma_{t\varphi}^\text{W}&= -\frac{\mu_0}{6} r_0 R \sin\Theta\left\{1+\frac{\mu_0}{30}\left(R^2+\frac{r_0^2}{2}\right)  \right.\nonumber
\\
&\quad{}+\left.\frac{\mu_0^2}{20} r_0^2\left[ \frac{R^2}{7}\left( \cos^2\Theta-\frac{103}{18}\right)-\frac{13}{3}\sigma_1  \right]\right\},
\\[1ex]
\gamma_{tt}^\text{W}&=-1-\frac{\mu_0}{6}R^2-\frac{\mu_0^2}{180}\left[ R^4 -2r_0^2R^2\left(1+2 \cos^2\Theta\right)\right] \nonumber
\\
&\quad{}+ \mu_0^3\frac{r_0^2}{90}\left[\sigma_1 R^2(2-\cos^2\Theta)+\frac{R^4}{14}\left(\frac{55}{6}-3\cos^2\Theta\right)
 \right]  \,.
\end{align}

After dealing with $\smash{\mu_0}$ and $r_0$, we have to find expressions for $b$ and $\kappa$. Recalling \cref{b-aprox}, we wrote $b^2$ as a series in $\smash{\mu_0}$ with coefficients $\sigma_1,\,\sigma_2$. To help with its determination, we can give more details about $\sigma_1$ and $\sigma_2$. When $b^2$ is written in terms of $\lambda$ and $\Omega$, its $\mathcal{O}(\lambda^0)$ terms will in general contain order $\Omega^2$ terms. These arise from $\mu_0r_0^2$ factors and, using dimensional arguments, we can redefine 
\begin{align}
 \sigma_1&\rightarrow \sigma_1r_s^2+r_0^2\nu_1\\
\sigma_2& \rightarrow(\sigma_2 r_s^2+r_0^2\nu_2)r_s^2
\end{align} 
to make this possibility more explicit during calculations.


Now we can write the approximate Wahl\-quist metric in terms of our parameters $\lambda$ and $\Omega$ using (\ref{constsubst}). Comparing the lower terms in $\lambda$ for $g_{t\varphi}$ of the CGMR co-rotating interior solution and the approximate Wahlquist metric just built, we can determine the proportionality constant $\kappa$ to be a series in our rotation parameter $\Omega$
\begin{equation}
\kappa= 1-\frac{\Omega^2}{10}+\mathcal{O}(\Omega^3)
\end{equation} 

\subsection{Adjusting terms}
Once  the relations between the approximation parameters of both metrics are determined   we can obtain the expression of the approximate Wahlquist metric written in the same parameters we have used for the CGMR co-rotating interior. With the coordinate change \labelcref{ttrans,wahl_rtrans,wahl_thetatrans} we eliminated terms that can not be present in CGMR. Now, to make both solutions coincide we can use changes of coordinates in the Wahlquist metric as long as they do not reintroduce undesired terms; also, we have freedom to adjust the $(m_0,\,m_2,\,j_1,\,j_3)$ constants of CGMR. Regarding the first, the remaining freedom is a change in the $\{r, \theta\} $ coordinates of the type displayed in \cref{change}. If we make this change in the Wahlquist metric
\begin{align}
r&\rightarrow r \left\{1+\lambda \Omega^2 \left(3 \sigma_1 \sin^2\theta-\frac12 \frac{r^2}{r_s^2} \cos^2\theta\right)\right.\nonumber
\\
&\quad \left.{}-\lambda^2 \left[\frac95\frac{r^2}{r_s^2} \sigma_1+\frac27\frac{r^4}{r_s^4}-\frac{1}{70} \Omega^2\frac{r^4}{r_s^4} \left(\frac{13}{3}+33 \cos^2\theta\right)\right]\right\}+\mathcal{O}(\lambda^3,\,\Omega^4),
\\[1ex]
\theta&\rightarrow \theta+\lambda \Omega^2 \sin\theta\cos\theta \left(\frac12\frac{r^2}{r_s^2}+3 \sigma_1-\frac{29}{210} \lambda \frac{r^4}{r_s^4}\right)+\mathcal{O}(\lambda^3,\,\Omega^4),
\end{align}
we get that, for the two metrics to be exactly equal up to $\mathcal{O}(\lambda^2,\,\Omega^3)$ the free constants (apart from $\lambda$ and $\Omega$) of the CGMR co-rotating interior must be
\begin{align}
m_0&=\mathcal{O}(\lambda^2,\,\Omega^4),& m_2 &=\frac65(1+2\lambda S)+\mathcal{O}(\lambda^2,\,\Omega^2),\label{enganche1}\\
 j_1 &= \frac95\Omega^2 S+\mathcal{O}(\lambda,\,\Omega^4),& j_3 &=\frac{36}{175}+\mathcal{O}(\lambda,\,\Omega^2) \label{enganche2}
\end{align}
and the free constants of the approximate Wahlquist metric must be
\begin{equation}
\sigma_1=S,\quad \sigma_2=0, \quad\nu_1=\frac12, \quad\nu_2=\frac1{18}.
\end{equation} 
This  gives $b^2$ as
\begin{equation}
b^2=(1+\Omega^2)(1+2\lambda S)+\mathcal{O}(\lambda^2,\,\Omega^4),
\label{wahl_bexp}
\end{equation} 
thus coinciding with the expansion of $\smash{\psi_\Sigma^2}$ from \labelcref{wahl_Sdef} if we take into account that the term $(1+\Omega^2)$ comes from the change of the normalization factor over the transformation of the temporal coordinate \cref{ttrans} we have done. This gives a first check of the consistency of the comparison since $b$  is the Wahlquist counterpart of $\psi_\Sigma$.

The final expressions for the metric components of either Wahlquist's solution or the CGMR interior in the orthonormal basis are, up to $\mathcal{O}(\lambda^2,\,\Omega^2)$ ---and $\mathcal{O}(\lambda^{3/2},\,\Omega^3)$ in $\gamma_{t\varphi}$---,
\begin{align}
\gamma_{rr}&=1- \lambda\frac{r^2}{r_s^2}\left\{1-\frac65\Omega^2P_2\right\}+ \frac{2}{5}\lambda^2\left\{-12S-\frac17 \frac{r^2}{r_s^2}\right.\nonumber
\\
&\quad{}+ \left.\Omega^2\left[\frac87\frac{r^2}{r_s^2}+\left(6S-\frac53\frac{r^2}{r_s^2}\right)P_2\right] \right\}\frac{r^2}{r_s^2},
\label{wahlfinal1}
\\
\gamma_{r\theta}&=\frac{31}{315}\lambda^2\Omega^2 \frac{r^4}{r_s^4}P_2^1,
\\[1ex]
\gamma_{\theta\theta}&=1- \lambda\frac{r^2}{r_s^2}\left(1-\frac65\Omega^2P_2\right)-\frac{2}{21}\lambda^2\frac{r^2}{r_s^2}\left\{ \frac{189}{5}S-\frac{12}{5} \frac{r^2}{r_s^2}\right.\nonumber
\\
&\quad{}+ \left.\Omega^2\left[-\frac{25}{6}\frac{r^2}{r_s^2}+\frac15\left(\frac{181}{3}\frac{r^2}{r_s^2}-126S\right)P_2\right] \right\},
\\[1ex]
\gamma_{\varphi\varphi}&=1- \lambda\frac{r^2}{r_s^2}\left(1-\frac65\Omega^2P_2\right)-\frac{2}{105}\lambda^2\frac{r^2}{r_s^2}\left\{ 189S-12 \frac{r^2}{r_s^2}\right.\nonumber
\\
&\quad{}+ \left.\Omega^2\left[\frac{17}{6}\frac{r^2}{r_s^2}+\left(\frac{110}{3}\frac{r^2}{r_s^2}-126S\right)P_2\right] \right\},
\\[1ex]
\gamma_{t\varphi}&=-\lambda^{1/2}\Omega\frac{r}{r_s}\left\{P_1^1+ \frac{\lambda}{5}\left[\frac{r^2}{r_s^2}P_1^1-3\Omega^2\left(\left(3S+\frac25\frac{r^2}{r_s^2}\right)P_1^1-\frac{2}{35}P_3^1\vphantom{\frac{{A^A}^A}{A^A}}\right)\right]\right\},
\\[1ex]
\gamma_{tt}&=-1- \lambda\frac{r^2}{r_s^2}\left[1-\frac23\Omega^2\left(1+\frac45P_2\right)\right]\nonumber
\\
&\quad{}+ \frac{\lambda^2}{5}\frac{r^2}{r_s^2}\left\{-\frac{r^2}{r_s^2}+\Omega^2\left[\frac83\frac{r^2}{r_s^2}+\left(12S+\frac{34}{21}\frac{r^2}{r_s^2}\right)P_2\right]\right\}.
\label{wahlfinal6}
\end{align}

To give another check of the whole procedure we can compare now  with the conditions necessary for our  $n=-2$ approximate metric to be of type Petrov D \cite{cuchi2010getting, toni2008ags,Cuchi:2012nmREVTEX}, i.e., \cref{petrovconD}.
%
They are compatible with the values of the constants $m_2$ and $j_3$ we have just found in \labelcref{enganche1,enganche2}, as wished. Also, when matched with an asymptotically flat vacuum exterior, $m_2$,  $j_3$ and the rest of the metric free constants can only have the expressions we found in \cite{Cuchi:2012nmREVTEX}. Since the $n=-2$ fluid for a type D interior does not satisfy the matched expressions, we concluded then that it can not be the source of such exterior in accordance with previous works \cite{wahlquist1968isf,bradley2000wmc,sarnobat2006wes}. Nevertheless, it is worth noting here that CGMR contains a $n=-2$ sub-case that lacks this problem and can indeed be matched that way. It has then all the characteristics of Wahlquist's fluid but Petrov type I instead of D.

Note, finally, that the Cartesian coordinates associated  to the spherical-like coordinates used above are not harmonic. Nevertheless, since  \crefrange{wahlfinal1}{wahlfinal6} correspond as well to the co-rotating $n=-2$ CGMR interior with particular values of the free constants, undoing the change \labelcref{stopchange} they become harmonic again.

\section{Remarks}
In this work we have taken the singularity free Wahlquist metric and managed to transform it into the form the CGMR interior metric takes when written in a co-rotating coordinate system. We have started from a formal expansion of Wahlquist's solution in $(\mu_0,\,r_0)$ and found its expression in terms of the parameters of CGMR, so it possesses the range of applicability already discussed for CGMR.

We have identified  Wahlquist's parameters corresponding to $\lambda$ and $\Omega$ of \cite{cabezas2006ags}. Doing this, we have found an expansion of the parameter $r_0$ of Wahlquist's metric  in terms of our $\Omega$. Accordingly, now we have an approximate expression of $r_0$ in terms of the better characterised quantities $\omega$ and $\mu_0$
\begin{equation}
 r_0= -\frac{r_s}{\sqrt{\lambda}}\Omega\left(1-\frac{\Omega^2}{10}\right)+\mathcal{O}(\Omega^4)= -\frac{6}{\mu_0}\omega\left(1-\frac{3 \omega^ 2}{5 \mu_0}\right)+\mathcal{O}\left(\frac{\omega^4}{\mu_0^2}\right).
\label{r0andomega}
\end{equation} 
To the best of our knowledge its  qualitative relation with the angular velocity was previously only guessed through the singular limiting procedure that takes the Wahlquist solution and leads to Whittaker's metric but no parametrization of it in terms of well defined quantities had been given. 

In the context of fixed EOS, this last equation, together with \cref{wahl_bexp}, completes the map from the free parameters of Wahlquist's solution  $(\,b,\, r_0)$ to the free parameters of a particular CGMR metric $(r_s,\,\omega)$. Curiously, we have gained insight in both sets. The role of $r_0$ as key to a vanishing twist vector and its good behaviour in the comparison with $\Omega$ shows far more clearly than the limiting procedure \cref{staticlimit} its relation with the rotation in the Wahlquist metric. But also, the role of $b$ as fundamental parameter in Wahlquist's solution hints towards the possibility of trying to build our post-Minkowskian approximation with a stronger emphasis on $\psi_\Sigma$ instead of the coordinate dependent $r_s$.

Last, notice that the usual interpretation of $\omega={u_\varphi}/{u_t}$ as angular velocity of the fluid as seen from the infinite lacks sense if we deal with a metric that is not matched with an asymptotically flat exterior. In our interior though, it is still singled out by the harmonic coordinate condition. Besides, the definition of stationarity and axisymmetry allows a change of coordinates $\{t=t',\,\varphi=\varphi'+at'\}$ that can modify the value of $\omega$ to 
$\omega'={u_\varphi'}/{u_t'}=\omega-a$ or make it zero (the case of co-rotating frames). Nevertheless, when dealing with a family of metrics explicitly dependent on $\omega$, its value can be important.  In the case of, e.\,g., CGMR, we see that written in co-rotating coordinates $u_{t'}/u_{\varphi'}=0$ but $\omega$ is part of the metric functions and actually, $\omega\rightarrow 0$ still leads to a static metric. It is actually the only way for the module of the CGMR twist vector 
\begin{equation}
 \varpi^\text{CGMR}=\frac{2 \lambda^{1/2}  \Omega }{r_s}+\mathcal{O}(\lambda^{3/2},\,\Omega^3)=2\omega+\mathcal{O}(\lambda^{3/2},\,\Omega^3)
\end{equation} 
to vanish (its $\mathcal{O}(\lambda^{3/2},\,\Omega^3)$ terms are proportional to $\omega$ as well). In this sense, the characterization of $r_0$ \labelcref{r0andomega} is meaningful.



\begin{acknowledgments}
We are very grateful to our reviewers for some important improvements on the original manuscript. This work was supported by the Spanish government grants FIS2006-05319, FIS2007-63034, FIS2009-07238 and FIS2012-30926. JEC thanks Junta de Castilla y Le\'on for PhD grant EDU/1165/2007.
\end{acknowledgments}

\bibliography{bibliografMACROsepnames}
\end{document}